\documentstyle[12pt,bezier]{article}
\begin{document}
% ***************    NEW COMMANDS   *******************
\def \inbar{\vrule height1.5ex width.4pt depth0pt}
\def \xC{\relax\hbox{\kern.25em$\inbar\kern-.3em{\rm C}$}}
\def \xR{\relax{\rm I\kern-.18em R}}
\newcommand{\xZ}{Z \hspace{-.08in}Z}
\newcommand{\xbe}{\begin{equation}}
\newcommand{\be}{\xbe}
\newcommand{\xee}{\end{equation}}
\newcommand{\ee}{\xee}
\newcommand{\xbea}{\begin{eqnarray}}
\newcommand{\bea}{\xbea}
\newcommand{\xeea}{\end{eqnarray}}
\newcommand{\eea}{\xeea}
\newcommand{\xnn}{\nonumber}
\newcommand{\nn}{\xnn}
\newcommand{\xkt}{\rangle}
\newcommand{\kt}{\xkt}
\newcommand{\xbr}{\langle}
\newcommand{\br}{\xbr}
\newcommand{\xlll}{\left( }
\newcommand{\xrrr}{\right)}
\newcommand{\xcun}{\mbox{\footnotesize${\cal N}$}}
\newcommand{\cun}{\mbox{\footnotesize${\cal N}$}}
\title{A New Class of Adiabatic Cyclic States and Geometric 
Phases for Non-Hermitian Hamiltonians}
\author{Ali Mostafazadeh\thanks{E-mail address: 
amostafazadeh@ku.edu.tr}\\ \\
Department of Mathematics, Ko\c{c} University,\\
Istinye 80860, Istanbul, TURKEY}
\date{ }
\maketitle

\begin{abstract}
For a $T$-periodic non-Hermitian Hamiltonian $H(t)$, we construct a class 
of adiabatic cyclic states of period $T$ which are not eigenstates of the
initial Hamiltonian $H(0)$. We show that the corresponding adiabatic geometric
phase angles are real and discuss their relationship with the conventional
complex adiabatic geometric phase angles. We present a detailed calculation
of the new adiabatic cyclic states and their geometric phases for a 
non-Hermitian analog of the spin 1/2 particle in a precessing magnetic field.  
\end{abstract}
\vspace{3mm}
%PACS numbers: 03.65.Bz
%\vspace{3mm}

\baselineskip=24pt

Since the publication of Berry's paper \cite{berry1984} on the adiabatic 
geometrical phase, the subject has undergone a rapid development. Berry's 
adiabatic geometric phase for periodic Hermitian Hamiltonians with a discrete
nondegenerate spectrum has been generalized to arbitrary changes of a quantum
state. In particular, the conditions on the adiabaticity \cite{aa} and cyclicity 
\cite{cyclic,po-sj,p31} of the evolution, Hermiticity of the Hamiltonian 
\cite{ga-wr}, and degeneracy \cite{wi-ze} and discreteness \cite{newton} of 
its spectrum have been lifted. Moreover, the classical \cite{hannay} and 
relativistic \cite{ga-ca} analogues of the geometric phase have been considered.

The purpose of this note is to offer an alternative generalization of Berry's 
treatment of the adiabatic geometric phase for a non-Hermitian parametric 
Hamiltonian $H[R]$ with a nondegenerate discrete spectrum. The parameters 
$R=(R^1,\cdots,R^n)$ are assumed to be real coordinates of a smooth 
parameter manifold, with the eigenvalues $E_n[R]$ and eigenvectors 
$|\psi_n,R\xkt$ of $H[R]$ depending smoothly on $R$. The approach 
pursued in the present paper differs from those of 
Refs.~\cite{ga-wr,da-mi-to,mi-si-ba-be,mo-he} in the choice of the 
adiabatically evolving state vectors. The corresponding geometric phase angle, 
which is shown to be real, has the same expression as the geometric phase angle
for a Hermitian Hamiltonian. It differs from the conventional complex 
geometric phase angle by terms which are small for an adiabatically
evolving system. In particular, this implies that in a generic adiabatic
evolution the imaginary part of the complex adiabatic  geometric
phase angle is small.

The generalization of the results of Berry \cite{berry1984} to non-Hermitian 
Hamiltonians was originally considered by Garrison and Wright \cite{ga-wr} 
and further developed by Dattoli et al \cite{da-mi-to}, Miniatura et al 
\cite{mi-si-ba-be}, and Mondrag\'on and Hernandez \cite{mo-he}. The main 
ingredient used by all of these authors in their derivation of  a non-Hermitian 
analog of Berry's phase is the biorthonormal eigenbasis of the Hamiltonian. 
More specifically, one writes the evolving state vector $|\psi(t)\xkt$ in an 
eigenbasis $|\psi_n,t\xkt:=|\psi_n,R(t)\xkt$ of the Hamiltonian $H(t):=H[R(t)]$,
	\xbe
	|\psi(t)\xkt=\sum_n C_n(t) |\psi_n,t\xkt\;,
	\label{expansion-1}
	\xee
and enforces the Schr\"odinger equation
	\xbe
	i|\dot\psi(t)\xkt=H(t)|\psi(t)\xkt\;,
	\label{sch-eq}
	\xee
where $R(t)$ denotes the curve traced in the parameter space, $C_n(t)$ are 
complex coefficients, a dot denotes a time-derivative, and $\hbar$ is set to 1. 
The resulting equation together with the eigenvalue equation for the 
Hamiltonian,
	\xbe
	H[R]|\psi_n,R\xkt=E_n[R]|\psi_n,R\xkt\;
	\label{eg-va-eq}
	\xee
where $R=R(t)$, yield
	\xbe
	\sum_n\left[ iC_n(t)|\dot\psi_n,t\xkt+
	[i\dot C_n(t)-E_n(t)C_n(t)]|\psi_n,t\xkt
	\right]=0\;.
	\label{eq-1}
	\xee
Next one takes the inner product of both sides of Eq.~(\ref{eq-1}) with the 
eigenvectors $|\phi_n,t\xkt=|\phi_n,R(t)\xkt$ of 
$H^\dagger(t)=H^\dagger[R(t)]$ which are defined by
	\xbea
	&&
	H^\dagger[R]|\phi_n,R\xkt=E_n^*[R]|\phi_n,R\xkt\;,
	\label{eg-va-eq-phi}\\
	&&\xbr\phi_m,R|\psi_n,R\xkt=\delta_{mn}\;.
	\label{biortho}
	\xeea
One then finds
	\xbe
	iC_m(t)\xbr\phi_m,t|\dot\psi_m,t\xkt+
	i\dot C_m(t)-E_m(t)C_m(t)+\sum_{n\neq m}
	iC_n(t)\xbr\phi_m,t|\dot\psi_n,t\xkt=0\;.
	\label{eq-2}
	\xee

For an adiabatically changing Hamiltonian,
	\xbe
	\xbr\phi_m,t|\dot\psi_n,t\xkt=\frac{\xbr\phi_m,t|\dot 
	H(t)|\psi_n,t\xkt}{E_n(t)-E_m(t)}\;,~~~~{\rm for}~~~
 	m\neq n
	\label{ad-ap}
	\xee
are small and the sum in (\ref{eq-2}) may be neglected \cite{ga-wr,ne-ra}. In 
this case, Eqs.~(\ref{eq-2}) can be easily integrated to yield
	\xbea
	C_m(t)&\approx&C_m(0)e^{i[\delta_m(t)+\gamma_m(t)]}\;,
	\label{total}\\
	\delta_m(t)&:=&-\int_0^t E_m(t')dt'\;,
	\label{dyn}\\
	\gamma_m(t)&:=&\int_0^ti\xbr\phi_m,t'|\dot\psi_m,t'\xkt
	dt'\;.
	\label{geo}
	\xeea
The validity of the adiabatic approximation, i.e., Eq.~(\ref{total}), is measured
by the value of the adiabaticity parameter which is defined by
	\be
	\eta:=\frac{1}{\omega_0}\:{\rm Sup}_{n,m\neq n,t}
	(|\br\phi_m,t|\dot\psi_n,t\kt|)\;.
	\label{eta}
	\ee
Here `Sup' stands for `Supremum' and $\omega_0$ is the frequency (energy)
scale of the system, \cite{pra-97a}. Adiabatic approximation is valid if and
only if $\eta\ll 1$. It is exact if and only if $\eta=0$.

For a periodic Hamiltonian with period $T$, where $R(T)=R(0)$, the 
eigenvectors of the initial Hamiltonian undergo approximate cyclic evolutions,
i.e., for  $|\psi(0)\xkt=|\psi_n,0\xkt$,
	\xbe
	|\psi(T)\xkt\approx e^{i[\delta_m(T)+\gamma_m(T)]}
	|\psi_n,0\xkt\;.
	\label{cyc-ev}
	\xee
In this case $\delta_m(T)$ and $\gamma_m(T)$ are called the adiabatic 
dynamical and geometrical phase angles, respectively. The geometric phase 
angle $\gamma_m(T)$ which can also be expressed in the form
	\xbe
	\gamma_m(T)=\oint i\xbr\phi_m,R|d|\psi_m,R\xkt\;,
	\label{gamma}
	\xee
is the complex analogue of Berry's adiabatic geometrical phase angle 
\cite{berry1984}. In Eq.~(\ref{gamma}), $d$ stands for the exterior derivative
with respect to the parameters $R^i$.

The approximate equations (\ref{total}) and (\ref{cyc-ev}) tend to exact
equations only in the extreme adiabatic limit, $\eta=0$, where the eigenvectors
of the Hamiltonian become stationary and the adiabatic approximation is exact
\cite{bohm-qm}. In this case, however, the geometric phase angle 
(\ref{gamma}) vanishes identically. For a system with nonstationary energy 
eigenstates, $\eta\neq 0$, and the adiabatic approximation is never exact 
\cite{bohm-qm}. In this case the eigenstates of the initial Hamiltonian are only
approximately cyclic. Strictly speaking, the state represented by $|\psi(T)\kt$ 
lies in a neighborhood of the state represented by $|\psi(0)\kt$ whose radius
is of order $\eta$.\footnote{The states belong to the projective Hilbert
space which is endowed with the Fubini-Study metric or its infinite-dimensional
analog. The latter is used to define the notion of a neighborhood of a state.}

The main motivation for the present analysis is the fact that although (due to the
adiabaticity of the evolution) the components of $|\dot\psi_m,t\xkt$ along the 
`normal' directions to $|\psi_n,t\xkt$ are negligible, $|\dot\psi_m,t\xkt$ does 
have non-negligible components along other eigenvectors. This is mainly 
because for a non-Hermitian Hamiltonian $|\psi_m,t\xkt$ do not form an 
orthogonal basis. This observation suggests an alternative expansion for the 
evolving state $|\psi(t)\xkt$, namely
	\xbe
	|\psi(t)\xkt=\sum_{n\neq m}C_n(t)|\psi_n,t\xkt+
	\tilde C(t)|\phi_m,t\xkt
	=\tilde C(t)\left[\sum_{n\neq m}\tilde C_n(t)
	|\psi_n,t\xkt+|\phi_m,t\xkt\right]\;.
	\label{new-exp}
	\xee
Here $m$ is a fixed label, $C_n(t)$ and $\tilde C(t)$ are complex coefficients, 
$\tilde C_n:=C_n/\tilde C$, and $\tilde C(t)$ is assumed not to vanish.

Although the expansions (\ref{expansion-1}) and (\ref{new-exp}) are 
mathematically equivalent, the latter allows for the construction of a new
class of adiabatic cyclic states of period $T$ which are not eigenstates
of the initial Hamiltonian. It should be emphasized that these states 
also have approximate cyclic evolutions. More specifically after each complete 
period $T$ of the Hamiltonian, the corresponding final state lies in
a neighborhood of the initial state with the radius of the neighborhood
being of order $\eta$.

Substituting (\ref{new-exp}) in the Schr\"odinger equation (\ref{sch-eq}), 
taking the inner product of both sides of the resulting equation first with 
$|\phi_m,t\xkt$ and then with $|\psi_k,t\xkt$ for $k\neq m$, and enforcing the 
adiabaticity condition: $\eta\ll 1$ which means
	\xbe
	\xbr\phi_m,t|\dot\psi_n,t\xkt\approx 0\approx
	\xbr\psi_k,t|\dot\phi_m,t\xkt\;,~~~{\rm for}~~~
	n\neq m\neq k,
	\label{ad-condi}
	\xee
one obtains
	\xbea
	&&i\xbr\phi_m,t|\phi_m,t\xkt\dot{\tilde C}(t)+\left[
	-\xbr\phi_m,t|\phi_m,t\xkt E_m(t)+i\xbr\phi_m,t|
	\dot\phi_m,t\xkt\right]\tilde C(t)\approx 0\;,
	\label{beta}\\
	&&\sum_{n\neq m}
	\left\{\xbr\psi_k,t|\psi_n,t\xkt\left[ \dot{\tilde C}_n(t)+
	\left(iE_n(t)+\frac{\dot{\tilde C}(t)}{\tilde C(t)}\right)
	\tilde C_n\right]+\xbr\psi_k,t|\dot\psi_n,t\xkt \tilde C_n(t)\right\}
	\xnn\\
	&&\hspace{7cm}\approx -i\xbr\psi_k,t|H(t)|\phi_m,t\xkt\;,
	~~~~{\rm for}~~~k\neq n\;.
	\label{alpha}
	\xeea
Eq.~(\ref{beta}) can be immediately integrated to yield
	\xbea
	\tilde C(t)&\approx&\tilde C(0)e^{i[\tilde\delta_m(t)+
	\tilde\gamma_m(t)]}\;,
	\label{tilde-c}\\
	\tilde\delta_m(t)&:=&-\int_0^t E_m(t')dt'\;,
	\label{tilde-delta}\\
	\tilde\gamma_m(t)&:=&\int_0^t\frac{i\xbr\phi_m,t'|
	\dot\phi_m,t'\xkt}{\xbr\phi_m,t'|\phi_m,t'\xkt}dt'\;.
	\label{tilde-gamma}
	\xeea
Using Eq.~(\ref{beta}), one can write Eq.~(\ref{alpha}) in the form:
	\xbea
	&&
	\sum_{n\neq m}
	\left\{\xbr\psi_k,t|\psi_n,t\xkt\left[\dot{\tilde C}_n(t)+
	i\left(E_n(t)-E_m(t)+\frac{i\xbr\phi_m,t|\dot\phi_m,t\xkt}{
	\xbr\phi_m,t|\phi_m,t\xkt}\right)\tilde C_n(t)\right]+
	\xbr\psi_k,t|\dot\psi_n,t\xkt \tilde C_n(t)\right\}\xnn\\
	&&\hspace{4cm}\approx -i\xbr\psi_k,t|H(t)|\phi_m,t\xkt~~~~~
	{\rm for}~~~k\neq n\;.
	\label{alpha'}
	\xeea
This is a system of first order linear ordinary differential equations with 
$T$-periodic coefficients. If the initial conditions are such that its solution is 
$T$-periodic, i.e., $\tilde C_n(T)\approx\tilde C_n(0)$, then the evolving state
undergoes an approximate cyclic evolution,
	\xbe
	|\psi(T)\xkt\approx e^{i[\tilde\delta_m(T)+\tilde\gamma_m(T)]}
	|\psi(0)\xkt\;.
	\label{cyc-tilde}
	\xee

As seen from Eq.~(\ref{cyc-tilde}), the dynamical part of the total (complex) 
phase angle, namely $\tilde\delta_m(T)$ has the same form as (\ref{dyn}), but 
the geometric part which can be written in the form
	\xbe
	\tilde\gamma_m(T)=\oint \frac{i\xbr\phi_m,R|
	d\phi_m,R\xkt}{\xbr\phi_m,R|\phi_m,R\xkt}\;,
	\label{new}
	\xee
differs from (\ref{gamma}). An interesting property of this geometric phase 
angle is that it has the same form as the geometric phase angle for a Hermitian
Hamiltonian, \cite{berry1984}. It is also very easy to show that 
$\tilde\gamma_m(T)$ is real.

We wish to emphasize that the (cyclic) geometric phase obtained here is a 
physical quantity, if  Eq.~(\ref{alpha'}) has a periodic solution. Although this is
a linear first order system with $T$-periodic coefficients, its solutions are not
generally $T$-periodic \cite{ince}. In the following, we shall first elaborate
on the relationship between the conventional complex geometric phase angle 
(\ref{gamma}) and our real geometric phase angle (\ref{tilde-gamma}). We shall
then explore the condition of the existence of periodic solutions of (\ref{alpha'})
for the simplest nontrivial case, i.e., the two-level system.

We first introduce
	\bea
	&&{\cal A}_{mn}(t):=i\br\phi_m,t|\dot\psi_n,t\kt\;,~~~
	\tilde{\cal A}_{mn}(t):=\frac{i\br\phi_m,t|\dot\phi_n,t\kt}{
	\br \phi_n,t|\phi_n,t\kt}\;,
	\label{aaaa}\\
	&&{\cal A}_m(t):={\cal A}_{mm}(t)\;,~~~
	\tilde{\cal A}_m(t):=\tilde{\cal A}_{mm}(t)\;,
	\label{aaaa2}
	\eea
so that
	\[\gamma_m(t)=\int_0^t {\cal A}_m(t')dt'\;,~~~~{\rm and}~~~~
	\tilde\gamma_m(t)=\int_0^t\tilde{\cal A}_m(t')dt'\;.\]
Now using the completeness of the biorthobormal basis vector, i.e.,
$\sum_{n}|\psi_n,t\kt\br\phi_n,t|=1$, we can write $|\phi_m,t\kt=
\sum_n\br\phi_n,t|\phi_m,t\kt~|\psi_n,t\kt$. Substituting this relation in the
definition of $\tilde{\cal A}_m(t)$ and performing the necessary algebra, we
obtain
	\be
	\tilde{\cal A}_m(t)={\cal A}_m(t)+
	i\frac{d}{dt}\,\ln \br\phi_m,t|\phi_m,t\kt
	+\sum_{n\neq m}\left(\frac{\br\phi_n,t|\phi_m,t\kt}{
	\br\phi_m,t|\phi_m,t\kt}\: {\cal A}_{mn}(t)\right)\;.
	\label{A=A}
	\ee
Therefore, the complex and real geometric phase angles are related by
	\be
	\tilde\gamma_m(T)=\gamma_m(T) +\sum_{n\neq m}\int_0^T
	\frac{\br\phi_n,t'|\phi_m,t'\kt {\cal A}_{mn}(t')\:dt'
	}{\br\phi_m,t'|\phi_m,t'\kt}\;.
	\label{G=G}
	\ee
Next we observe that according to Eqs.~(\ref{eta}), (\ref{aaaa}), and 
(\ref{A=A}) the difference between  $\tilde{\cal A}_m$ and ${\cal A}_m$
consists of terms which are bounded by $\eta$. Therefore, in the extreme
adiabatic limit $\eta\to 0$, the two phase angles coincide. This in turn
shows that in this limit the imaginary part of the complex geometric 
phase angle $\gamma_m(T)$ tend to zero. In practice, however, $\eta$ has
a small but nonzero value (unless in the trivial case where the energy 
eigenvectors are stationary and $\gamma_m(T)=0=\tilde\gamma_m(T)$). In
this case, $\gamma_m(T)\neq\tilde\gamma_m(T)$. This is consistent with 
the fact that $\gamma_m(T)$ is not necessarily real.

In the remainder of this paper we shall apply our general results to study the
two-level system. This system has been the subject of detailed study 
\cite{da-to-mi,kv-pu,mo-he} for its physical applications in particular in 
connection with the spontaneous decay of the excited states of atoms 
\cite{kv-pu}.

Consider a two-dimensional Hilbert space where the Hamiltonian is a possibly
non-Hermitian $2\times 2$ complex matrix with distinct eigenvalues. Setting 
$m=2$ in Eqs.~(\ref{new-exp}) and using Eqs.~(\ref{alpha'}), one has
	\xbea
	&&|\psi(t)\xkt=\tilde C(t)\left[\tilde C_1(t)|\psi_1,t\xkt
	+|\phi_2,t\xkt\right]\;,
	\label{new-exp-2}\\
	&&\dot{\tilde C}_1+Q(t)\tilde C_1={\cal R}(t)\;,
	\label{eq-3}
	\xeea
where
	\xbea
	Q(t)&:=&i[E_1(t)-E_2(t)]+
	\frac{\xbr\psi_1,t|\dot\psi_1,t\xkt}{\xbr\psi_1,t|
	\psi_1,t\xkt}-\frac{\xbr\phi_2,t|\dot\phi_2,t\xkt}{
	\xbr\phi_2,t|\phi_2,t\xkt}\;,\xnn\\
	{\cal R}(t)&:=& \frac{-i\xbr\psi_1,t|H(t)|\phi_2,t\xkt}{
	\xbr\psi_1,t|\psi_1,t\xkt}\;.\xnn
	\xeea
In Eqs.~(\ref{eq-3}) and the remainder of this paper, the adiabatic 
approximation is assumed to be valid and $\approx$'s are replaced
by  $=$'s.

Eq.~(\ref{eq-3}) can be easily integrated to yield
	\xbe
	\tilde C_1(t)=W(t)\left(\tilde C_1(0)+
	\int_0^t\frac{{\cal R}(t')}{W(t')}dt'\right)\;,
	\label{c=}
	\xee
where
	\xbe
	W(t):=e^{-\int_0^tQ(s)ds}\;.
	\label{W}
	\xee

Having obtained the general solution, one can easily check for the periodic 
solutions. In view of the fact that $Q$ and ${\cal R}$ are periodic functions of 
time with the same period $T$ as the Hamiltonian, one can show that
	\xbe
	\tilde C_1(t+T)-\tilde C_1(t)=W(t)
	\left(\tilde C_1(0)[W(T)-1]+
	W(T)\int_0^T\frac{{\cal R}(t')}{W(t')}dt'\right)\;.
	\label{c=2}
	\xee
Therefore, the initial condition leading to a periodic solution is given by
	\xbe
	\tilde C_1(0)=\frac{W(T)\int_0^T\frac{{\cal R}(t')}{W(t')}dt'}{
	1-W(T)}\;.
	\label{cyc-condi-1}
	\xee
If both the numerator and denominator on the right hand side of 
(\ref{cyc-condi-1}) vanish, then the right hand side of (\ref{c=2}) vanishes and
the solution (\ref{c=}) is always periodic. If the denominator vanishes, i.e.,
	\xbe
	W(T):=e^{i\int_0^T[E_2(s)-E_1(s)]ds}
	e^{-i\oint\left(\frac{i\xbr\phi_2,R|
	d\phi_2,R\xkt}{\xbr\phi_2,R|\phi_2,R\xkt}-
	\frac{i\xbr\psi_1,R|d\psi_1,R\xkt}{
	\xbr\psi_1,R|\psi_1,R\xkt}\right)}= 1 \;,
	\label{cyc-condi-2}
	\xee
but the numerator does not, there is no periodic solutions. If the denominator 
does not vanish, $W(T)\neq 1$, then there is a particular initial condition given 
by Eq.~(\ref{cyc-condi-1}) that leads to a periodic solution. It is for this 
solution that the formula (\ref{tilde-gamma}) for the (cyclic) adiabatic geometric
phase applies.

Clearly a similar treatment may be carried out for the choice $m=1$ in 
Eq.~(\ref{new-exp}). The corresponding results are given by the same formulas
as for the case $m=2$ except that one must interchange the labels 1 and 2.

For the case of a Hermitian Hamiltonian where $|\phi_n,R\xkt=|\psi_n,R\xkt$, 
the right hand side of Eq.~(\ref{alpha'}) vanishes and the trivial solution: 
$\tilde C_n=0$ is periodic. This solution corresponds to the well-known result 
that for an adiabatically changing Hamiltonian the eigenvectors $|\psi_n,0\xkt$
of the initial Hamiltonian undergo cyclic evolutions. In this case, 
Eq.~(\ref{tilde-gamma}) gives Berry's phase \cite{berry1984}.

Let us next consider the two-level system with the parametric Hamiltonian
	\xbe
	H=H[E,\theta,\varphi]:=E\left(\begin{array}{cc}
	\cos\theta & e^{-i\varphi} \sin\theta \\
	e^{i\varphi} \sin\theta & -\cos\theta
	\end{array}\right)\;,
	\label{H}
	\xee
where $\theta\in\xR$ and $E,\varphi\in\xC$, \cite{kv-pu}. Note that due to the
form of the Hamiltonian (\ref{H}), $\theta\in[0,\pi]$ and the real part 
$\varphi_r$ of $\varphi$ has the range $[0,2\pi)$, whereas $E$ and the 
imaginary part $\varphi_i$ of $\varphi$ can take arbitrary complex and real 
values, respectively.

It is not difficult to show that the eigenvalues of $H$ are given by $\pm E$. 
Hence for $E\neq 0$ one has two distinct eigenvalues: $E_1=-E$ and 
$E_2=E$. The corresponding eigenvectors of $H$ and $H^\dagger$ are given 
by
	\xbea
	&&|\psi_1\xkt=\left(\begin{array}{c}
	-e^{-i\varphi}\sin({\theta\over 2})\\
	\cos({\theta\over 2})\end{array}\right)\;,~~~~
	|\psi_2\xkt=\left(\begin{array}{c}
	\cos({\theta\over 2})\\
	e^{i\varphi}\sin({\theta\over 2})\end{array}\right)\;,~~~{\rm and}
	\label{psi}\\
	&&|\phi_1\xkt=\left(\begin{array}{c}
	-e^{-i\varphi^*}\sin({\theta\over 2})\\
	\cos({\theta\over 2})\end{array}\right)\;,~~~~
	|\phi_2\xkt=\left(\begin{array}{c}
	\cos(\frac{\theta}{2})\\
	e^{i\varphi^*}\sin({\theta\over 2})\end{array}\right)\;,
	\label{phi}
	\xeea
respectively. The geometric phase angles $\gamma_m(T)$ and 
$\tilde\gamma_m(T)$ can be easily computed,
	\xbea
	&&\gamma_1(T)=-\gamma_2(T)={1\over 2}\oint(1-\cos\theta)d\varphi\;,
	\label{gamma=}\\
	&&\tilde\gamma_1(T)=\oint\frac{d\varphi_r}{1+e^{2\varphi_i}\cot^2
	({\theta\over 2})}\;,~~~~
	\tilde\gamma_2(T)=-\oint\frac{d\varphi_r}{1+e^{-2\varphi_i}\cot^2
	({\theta\over 2})}\;.
	\label{tilde-gamma-2=}
	\xeea

Let us next consider the case where $E,~\theta$ and $\varphi_i$ are constant
and $\varphi_r=\omega t$. In this case, the Hamiltonian (\ref{H}) is a 
non-Hermitian analogue of the Hamiltonian of a magnetic dipole interacting 
with  a precessing magnetic field \cite{berry1984,bohm-qm,jmp97}. For this 
system, one can easily evaluate the integrals in (\ref{gamma=}) and 
(\ref{tilde-gamma-2=}), and obtain
	\xbea
	&&\gamma_1(T)=-\gamma_2(T)=\pi(1-\cos\theta)\xnn\\
	&&\tilde\gamma_1(T)=\frac{2\pi}{1+e^{2\varphi_i}
	\cot^2({\theta\over 2})}\;,~~~~
	\tilde\gamma_2(T)=\frac{-2\pi}{1+e^{-2\varphi_i}
	\cot^2({\theta\over 2})}\;,\xnn
	\xeea
where $T:=2\pi/\omega$.

Furthermore, one can obtain the explicit form of the initial condition for which
the state vector (\ref{new-exp-2}) performs an adiabatic cyclic evolution. Using 
Eqs.~(\ref{cyc-condi-1}), (\ref{psi}), and (\ref{phi}), one finds
	\xbe
	\tilde C_1(0)=-\frac{\sinh\varphi_i \sin\theta}{
	1+(\frac{\pi}{ET})\left(\frac{
	e^{2\varphi_i}-\cot^2(\theta/2)}{
	e^{2\varphi_i}+\cot^2(\theta/2)}\right)}\;.
	\xnn
	\xee

One can repeat the above analysis for the state vector 
	\[|\psi(t)\xkt=\tilde C(t)\left[\tilde C_2(t)|
	\psi_2,t\xkt+|\phi_1,t\xkt\right]\]
which corresponds to the choice $m=1$ in Eqs.~(14) -- (23). In this case the 
initial condition leading to an adiabatic cyclic evolution is given by
	\xbe
	\tilde C_2(0)=-\frac{\sinh\varphi_i \sin\theta}{
	1+(\frac{\pi}{ET})\left(\frac{
	e^{-2\varphi_i}-\cot^2(\theta/2)}{
	e^{-2\varphi_i}+\cot^2(\theta/2)}\right)}\;.
	\xnn
	\xee

It is not difficult to see that for $\varphi_i=0$ the above results tend to those of
Berry \cite{berry1984}. It is also remarkable that unlike $\gamma_m(T)$ the 
new geometric phase angles $\tilde\gamma_m(T)$ are sensitive to the 
imaginary part $\varphi_i$ of $\varphi$.

In conclusion, we wish to emphasize that the issue of the 
existence of (exact) cyclic states of arbitrary period for a general
Hermitian Hamiltonian is addressed in Ref.~\cite{anandan-88}. The
results of \cite{anandan-88} also generalize to the non-Hermitian
Hamiltonians. This is simply because the existence of cyclic states
of period $\tau$ is identical with the existence of eigenstates
of the evolution operator $U(\tau)$. In the present article, we have
considered the adiabatic cyclic states which have approximate 
cyclic evolutions. What we have established is the existence and
construction of a class of adiabatic cyclic states which have
the same period as the Hamiltonian but are not eigenstates of the
initial Hamiltonian. These states have the same cyclicity properties as 
the well-known eigenstates of the initial Hamiltonian. The only 
difference is that the new adiabatic cyclic states acquire real
geometric phase angles. For a two-level non-Hermitian Hamiltonian
(\ref{H}) we have constructed these states explicitly and computed the
corresponding geometric phases. For this system if we choose $E,~\theta$
and the imaginary part $\varphi_i$ of $\varphi$ to be constant and 
require the real part $\varphi_r$ of $\varphi$ to be proportional to
time, i.e., $\varphi_r(t)=\omega t$, then both the conventional
geometric phase angle $\gamma_m(T)$ and the geometric phase angle 
$\tilde\gamma_m(T)$ introduced in this paper are real. However, 
unlike $\gamma_m(T)$,  $\tilde\gamma_m(T)$ depend on $\varphi_i$.

Finally we wish to note that one may follow the approach of 
Refs.~\cite{po-sj,p31} to define a noncyclic analog of the adiabatic geometric
phase obtained in the present paper. In this case, one is not limited to the 
periodic solutions of Eq.~(\ref{alpha'}).

\end{document}